\documentstyle[aps,epsfig]{revtex}

\begin{document}
\title{A simple approximation for fluids with narrow \\ attractive potentials} 

\author{D.~Pini}
\address{Istituto Nazionale per la Fisica della Materia and Dipartimento 
di Fisica,
\\ Universit\`a di Milano, Via Celoria 16, 20133 Milano, Italy}

\author{A.~Parola}
\address{Istituto Nazionale per la Fisica della Materia and Dipartimento di 
Scienze Fisiche, \\
Universit\`a dell'Insubria, Via Valleggio 11, 22100 Como, Italy}

\author{L.~Reatto}
\address{Istituto Nazionale per la Fisica della Materia and Dipartimento 
di Fisica,
\\ Universit\`a di Milano, Via Celoria 16, 20133 Milano, Italy}

\maketitle

\begin{abstract}

We study a simple modification of the optimized random phase approximation
(ORPA) aimed at improving the performance of the theory for interactions 
with a narrow attractive well 
by taking into account contributions to the direct correlation
function non-linear in the interaction.
The theory is applied to a hard-core Yukawa and a square-well potential.
Results for the equation of state, the correlations, and the critical point
have been obtained for attractions of several ranges, and compared with 
Monte Carlo simulations. When the attractive interaction is narrow, 
the modified ORPA significantly improves over the plain one, especially
with regard to the consistency between different routes to the thermodynamics,
the two-body correlation function, and the critical temperature. 
However, while the spinodal curve of the modified theory is accessible, 
the liquid-vapor coexistence curve  
is not. A possible strategy to overcome this drawback is suggested. 

\end{abstract}

\section{The closure}

In the last years there has been a considerable interest in model fluids 
in which a hard-core or very steep repulsion in the interparticle potential
is followed by a narrow and deep attractive well. This interest mainly stems
from the fact that such potentials provide a modelization of the interaction
in many colloidal suspensions and protein solutions which, albeit crude,
succeeds nevertheless in capturing the most important features 
of thermodynamics~\cite{piazza}, structure~\cite{belloni}, 
and phase behavior~\cite{rosenbaum,lomakin} of these systems. 
When describing the properties of fluids with narrow attractive potentials,
it appears that the second virial coefficient plays an important role. 
In particular, while  
the critical temperature $T_{c}$ depends very sensitively on the interaction
range, the second virial coefficient $B_{2}$ evaluated at $T_{c}$ changes 
quite slowly~\cite{vliegenthart}. Remarkably, this remains true even when 
considering different interaction profiles, e.g. switching from a square-well 
to an attractive Yukawa potential. On the basis of simulation results, 
$B_{2}(T_{c})$ is found to lie in the neighborhood of 
$B_{2}(T_{c})\simeq -6s_{0}$~\cite{vliegenthart}, 
where $s_{0}=\pi\sigma^{3}/6$ is the volume 
of the particle, represented as a hard sphere of diameter $\sigma$.  
Other properties of fluids with short-range attractions exhibit a considerable
degree of universality when described in terms of $B_{2}$. For instance, 
the range of the potential itself can be defined as that  
of an ``equivalent'' square-well fluid with the same value of $B_{2}$ as  
the original interaction~\cite{noro}. With this prescription, 
the disappearance of a stable liquid-vapor transition characteristic 
of short-ranged attractive 
potentials~\cite{rosenbaum,lomakin,lomba,mederos,hagen} 
has been found to occurr at about the same range for several different 
interactions. It has also been observed that $B_{2}$ is a relevant parameter
in protein crystallization, the optimal conditions for crystallization 
corresponding to a narrow range of values of $B_{2}$~\cite{george}. 

Probably the most widespread method to model narrow attractive interactions
in colloidal systems makes use of Baxter's exact solution of the 
adhesive hard sphere (AHS) model~\cite{baxter} within the Percus-Yevick (PY) 
integral equation. As it is well known, this can be thought of 
as a square-well fluid in the limit of vanishing range and infinite well depth 
so that $B_{2}$ will stay
finite. Therfore, the relevance of the second virial coefficient in this model
stems from the fact that this is the quantity which allows one to make contact
with the real system. It is worthwhile recalling that the existence of 
an equilibrium solution for the AHS model is a consequence of the approximate
character of the PY theory. In fact, it has been pointed out~\cite{stell}
that the AHS model is in itself thermodynamically unstable. While 
Baxter's solution is successful in describing many properties
of systems with very short-ranged attractive potentials, one would also like
to describe these systems without resorting to the sticky limit. It is well
known that the most popular perturbative approaches in liquid-state theory,
such as the mean spherical approximation (MSA) and the optimized random phase
approximation (ORPA)~\cite{hansen} are not suited for short-ranged 
interactions. The reason
is that both of these theories assume that the direct correlation function 
$c(r)$ outside the repulsive core is simply proportional to the interaction
$w(r)$. This is a sound assumption whenever $w(r)$ is long-ranged, but becomes
less and less so as the range of $w(r)$ decreases to a small fraction 
of the size of the particles. An accurate description of the correlations
is provided by other integral-equation approaches, the most successful 
of which is probably the modified hypernetted chain (MHNC) 
theory~\cite{caccamorev}, in which
$c(r)$ depends on the interaction both directly and via the distribution
function $g(r)$. On the other hand, determining the critical point by these
more sophisticated methods proves difficult or even impossible, while this
is accomplished quite easily within the simpler perturbative schemes.
Therefore, it can be interesting to seek for a modification of the ORPA which
mantains the simplicity of the original formulation, while accounting more
accurately for the effects on $c(r)$ non-linear in the interaction.        
This could prove useful also in the context of more complex theories, namely
the renormalization-group based hierarchical reference theory (HRT)~\cite{hrt}
and the self-consistent Ornstein-Zernike approximation (SCOZA)~\cite{scoza},
both of which adopt a simple ORPA-like functional form for $c(r)$. 
These theories deliver the best results for the critical point and the 
liquid-vapor coexistence curve of simple fluids, but the inadequacy of ORPA
for very short-ranged interactions partly affects their accuracy 
in this regime, especially as far as the correlations 
are concerned~\cite{costa}. 

A possible way to modify the ORPA is suggested by the aforementioned 
observation that the second virial coefficient evaluated at the critical 
temperature depends little on the specific form of the interaction. 
For a potential $v(r)$ which is the sum of a hard-core contribution and 
an attractive tail $w(r)$, this behavior could be accounted for, 
at the mean-field level, by replacing the usual random phase approximation
(RPA) for the direct correlation function $c(r)$
\begin{equation}
c(r)=c_{\rm HS}(r)-\beta w(r)
\label{rpa}
\end{equation}
by the expression
\begin{equation}
c(r)=c_{\rm HS}(r)+e^{-\beta w(r)}-1 \, ,
\label{rpanl}
\end{equation}
where $\beta=1/k_{\rm B}T$ is the inverse temperature and $c_{\rm HS}(r)$ 
is the direct correlation function of the hard-sphere reference fluid. 
The latter can be considered as a known quantity, for instance by resorting
to the accurate Verlet-Weis parameterization~\cite{verlet}.
Eq.~(\ref{rpanl}) corresponds to a mean-field or van der Waals-like expression
for the Helmholtz free energy per unit volume $a$:
\begin{equation}
\beta a=\beta a_{\rm HS} + \frac{1}{2}\, \rho^{2}\int \!\! d^{3}{\bf r}
\, [1-e^{-\beta w(r)}] \, ,
\label{free}
\end{equation}
where $a_{\rm HS}$ is the Helmholtz free energy per unit volume 
of the hard-sphere reference fluid. As is readily seen from 
this expression, the mean-field critical density $\rho_{c}$ does not
depend on the attractive interaction $w(r)$ and is given by 
$\rho_{c}=0.249 \sigma^{-3}$ if the Carnahan-Starling expression~\cite{hansen} 
for the reduced compressibility of the hard-sphere gas 
$\chi^{\rm HS}_{\rm red}(\rho)$ is used. As a consequence, Eq.~(\ref{free})
gives a universal value for $B_{2}(T_{c})$ at the critical point:
\begin{equation}
B_{2}(T_{c})=\frac{1}{2}\int \!\! d^{3}{\bf r}\, [1-e^{-\beta_{c} v(r)}]=
\frac{2}{3}\pi \sigma^{3}-\frac{1}
{2\rho_{c}\chi^{\rm HS}_{\rm red}(\rho_{c})}=-6.601 \, s_{0} \, ,
\label{virial}
\end{equation}
where $\beta_{c}$ is the inverse critical temperature. This is not far from 
the above-mentioned value $B_{2}(T_{c})\simeq -6s_{0}$ reported in 
Ref.~\cite{vliegenthart}. We have then considered a similar modification
of the ORPA by replacing the perturbation $-\beta w(r)$ with the Mayer 
function $e^{-\beta w(r)}-1$. We will refer to this approximation as  
non-linear ORPA. The Ornstein-Zernike equation which relates 
the direct correlation function $c(r)$ to the radial distribution function
$g(r)$ is then closed by means of the following {\em ansatz}:
\begin{equation}
\left\{
\begin{array}{lc}
g(r)=0 & \mbox{\hspace{0.4cm}} r<\sigma \, , \\
c(r)=c_{\rm HS}(r)+e^{-\beta w(r)}-1  & \mbox{\hspace{0.4cm}} r>\sigma \, .
\end{array}
\right.
\label{orpanl}
\end{equation}     
This closure differs from Eq.~(\ref{rpanl}) insomuch as here, as in the usual
ORPA, we require $g(r)$ to vanish inside the core because of the singular
repulsion. Instead of Eq.~(\ref{rpanl}) we have then:
\begin{equation}
c(r)=c_{\rm HS}(r)+e^{-\beta w(r)}-1+{\cal G}(r) \, ,
\label{orpanl2}
\end{equation}
where the function ${\cal G}(r)$ vanishes for $r>\sigma$, and is determined 
so that the {\em core condition} $g(r)=0$ for $r<\sigma$ is satisfied. 
The latter can be written via the Ornstein-Zernike equation in terms 
of the direct correlation function in Fourier space $\hat{c}(k)$:
\begin{equation}
\int \!\! \frac{d^{3}{\bf k}}{(2\pi)^{3}}\, e^{i{\bf k}{\cdot}{\bf r}}
\frac{\hat{c}(k)}{1-\rho \hat{c}(k)}=-1, \mbox{\hspace{0.5cm}} r<\sigma \, .
\label{core}
\end{equation}
As is customary in ORPA calculations, we have implemented this condition 
in an approximate fashion by representing ${\cal G}(r)$ for $r<\sigma$ 
on a finite set of basis functions as a polynomial of given 
(typically, fourth) degree:
\begin{equation}
{\cal G}(r)=\sum_{j=1}^{n} u_{j}\, r^{j-1}, \mbox{\hspace{0.5cm}} r<\sigma \, .
\label{g}
\end{equation}
The state-dependent coefficients $u_{j}$, $j=1\ldots n$, are determined 
by projecting Eq.~(\ref{core}) on each basis function in the interval
$0<r<\sigma$. If we set:
\begin{equation}
P_{j}(r)=\left\{
\begin{array}{lc}
r^{j-1} & \mbox{\hspace{0.4cm}} r<\sigma \, , \\
0 & \mbox{\hspace{0.3cm}} r>\sigma 
\end{array}
\right.
\label{pol}
\end{equation}     
and we indicate by $f(r)$ the Mayer function $e^{-\beta w(r)}-1$, 
Eq.~(\ref{core}) gives the following set of $n$ non-linear equations 
for the $n$ coefficients $u_{j}$:
\begin{equation}
\int \!\! \frac{d^{3}{\bf k}}{(2\pi)^{3}} \, \hat{P}_{l}(k) 
\frac{\hat{c}_{\rm HS}(k)+\hat{f}(k)+\sum_{j=1}^{n}u_{j}\hat{P}_{j}(k)}
{1-\rho\left[\hat{c}_{\rm HS}(k)+\hat{f}(k)+\sum_{j=1}^{n}u_{j}\hat{P}_{j}(k)
\right]}=
-\hat{P}_{l}(k\!=\!0), \mbox{\hspace{0.5cm}} l=1\ldots n \, ,
\label{core2}
\end{equation}
where the hat has been used to denote Fourier transform. This system has 
been solved numerically by means of a standard Newton-Raphson iterative
algorithm. Particular care has been taken to ensure convergence 
of the algorithm even for states at which the compressibility of the system
is large, i.e. such that the denominator of the integrand in 
Eqs.~(\ref{core}), (\ref{core2}) becomes small near $k=0$. By evaluating 
analytically
the contribution to the integral at small wavevector $k$, we could easily 
achieve convergence for reduced compressibilities as large as $\sim 10^{8}$.
This enabled us to accurately locate the compressibility-route critical point 
and the spinodal curve
(that is, the locus of diverging compressibility) predicted by the theory.      
In order to check the reliability of the approximate procedure adopted 
to solve Eq.~(\ref{core}), we used the same method to solve the MSA 
for a hard-sphere fluid with an attractive Yukawa tail and compared the result
with the exact solution of the MSA for this particular potential~\cite{hoye}. 
The MSA is obtained from Eq.~(\ref{orpanl2}) by using for $c_{\rm HS}(r)$
the solution of the PY equation for the hard-sphere fluid and by replacing 
the Mayer function with the perturbation $-\beta w(r)$. The comparison 
shows that the results obtained by the procedure described above are nearly
undistinguishable from those of the exact solution for all the interaction
ranges investigated here. The relative error on the reduced compressibility 
$\chi_{\rm red}$ is typically of the order of $10^{-5}$ and does not exceed 
few percents for $\chi_{\rm red}$ in the range $10^{4}$--$10^{5}$, while
critical temperatures coincide to within four significant figures.   

\section{Results}

We have assessed the performance of the non-linear ORPA of Eq.~(\ref{orpanl})
by comparing its predictions for the thermodynamics, the correlations,
and the critical point with those of other liquid-state theories as well as 
simulations. Most of the results were obtained for a hard-core Yukawa 
fluid at several values of the interaction range. The potential has the 
form 
\begin{equation}
v(r)=\left\{
\begin{array}{ll}
\infty   &  r<\sigma \, ,  \\
-\epsilon \sigma \, e^{-z (r-\sigma)}/r
  \mbox{\hspace{0.3cm}} & r>\sigma \, ,
\end{array}
\right.
\label{yuk}
\end{equation} 
where $z$ is the inverse-range parameter. We also performed some calculations
for a hard-core square-well potential
\begin{equation}
v(r)=\left\{
\begin{array}{cl}
\infty   &  r<\sigma \, , \\
-\epsilon \mbox{\hspace{0.3cm}} & \sigma<r<(1+\delta)\, \sigma \, , \\
\mbox{\hspace{0.1cm}} 0 & r>(1+\delta)\, \sigma \, ,
\end{array}
\right.
\label{sq}
\end{equation}     
where $\delta$ measures the width of the attractive well. 
It is worthwhile observing that for this particular potential 
the closure~(\ref{orpanl}) 
amounts to evaluating the plain ORPA at an ``effective'' inverse temperature
$\beta_{\rm eff}$ given by $\beta_{\rm eff}\epsilon =e^{\beta\epsilon}-1$.
This corresponds to rescaling the amplitude of the perturbation $-\beta w(r)$
so that the second virial coefficient is correctly reproduced.
Such a prescription was in fact adopted several years ago
to improve the performance of the MSA for the very narrow square-well 
potentials used to model the structure factor of microemulsions~\cite{huang}. 
We also recall that the above interactions can be used to mimic 
a Lennard-Jones (LJ) fluid by suitably adjusting their ranges. 
Adequate values for this purpose are $z=1.8 \sigma^{-1}$~\cite{henderson} 
for the hard-core Yukawa potential~(\ref{yuk}) and $\delta=0.5$ 
for the square-well potential~(\ref{sq}). In the following 
we will deal with substantially shorter-ranged interactions.  

\begin{figure}
\centerline{\psfig{file=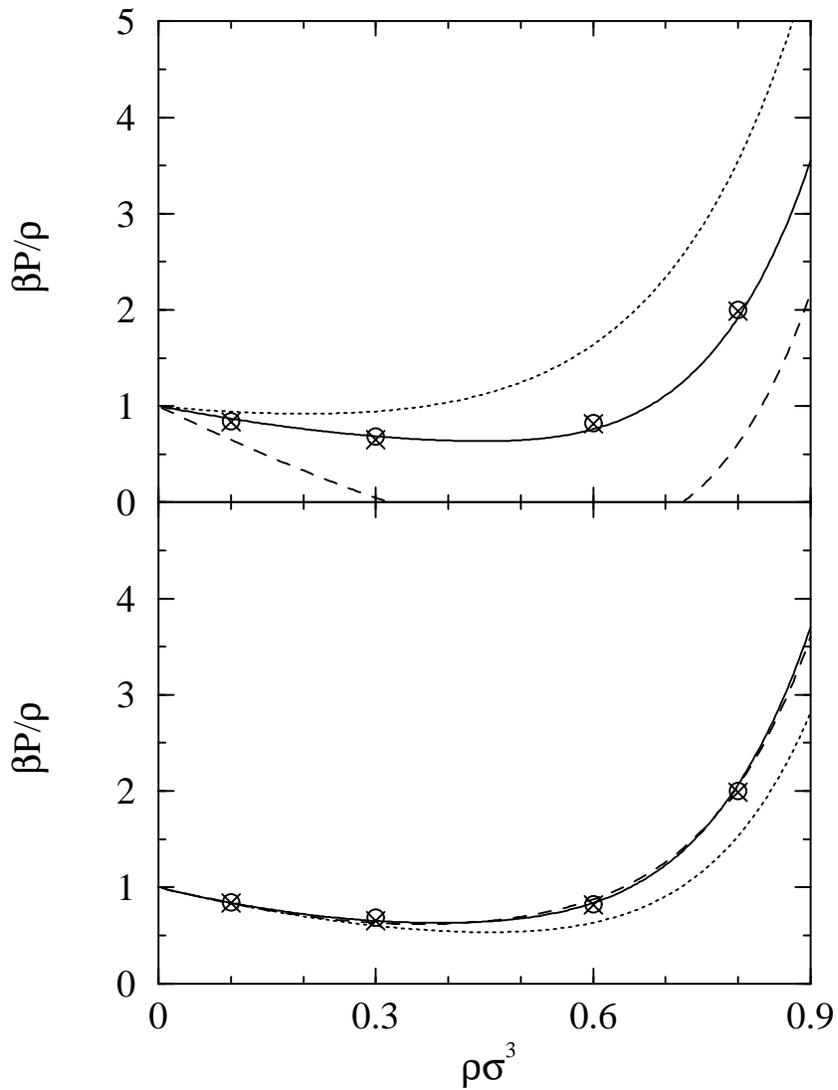}}
\caption{Compressibility factor of the hard-core Yukawa fluid 
(see Eq.~(\protect\ref{yuk}))
with inverse range $z=4\sigma^{-1}$ at reduced temperature 
$T^{*}=0.7$, corresponding
to $T/T_{c}\simeq 1.2$.
The compressibility, virial, and energy routes of the linear ORPA 
(upper panel) and non-linear
ORPA (lower panel) are compared with MHNC and MC results.
Upper panel: dotted line, linear ORPA compressibility route; dashed line,
linear ORPA virial route; solid line: linear ORPA energy route; crosses, MHNC;
and circles, MC simulation results~\protect\cite{shukla}. Lower panel:
same notation as in the upper panel for the non-linear ORPA.}  
\label{fig:zcomp1} 
\end{figure}

\begin{figure}
\centerline{\psfig{file=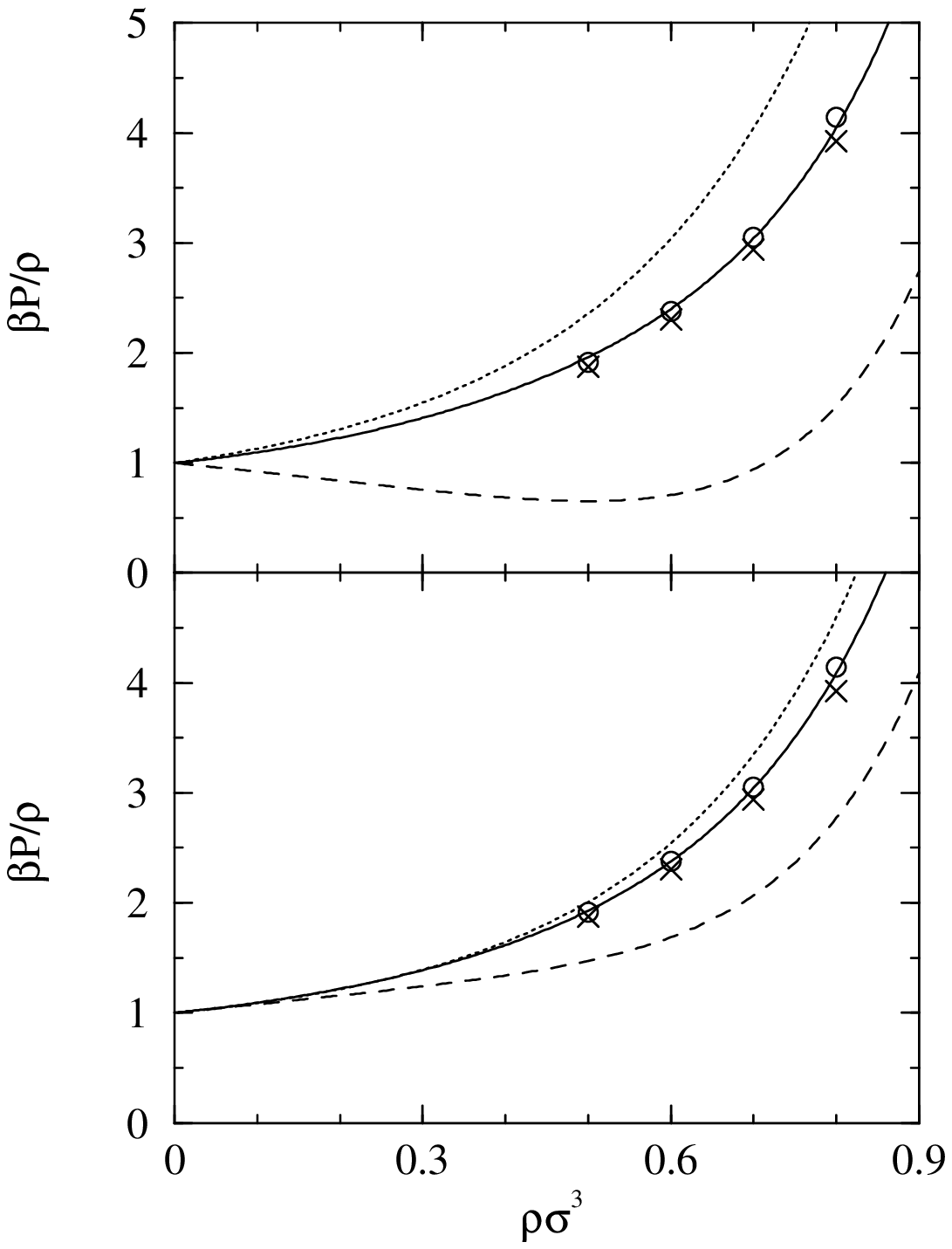}} 
\caption{Same as figure~\protect\ref{fig:zcomp1} with $z=9\sigma^{-1}$ 
and $T^{*}=0.8$. MHNC and MC results are from Ref.~\protect\cite{costa}.}
\label{fig:zcomp2}
\end{figure}

\begin{figure}
\centerline{\psfig{file=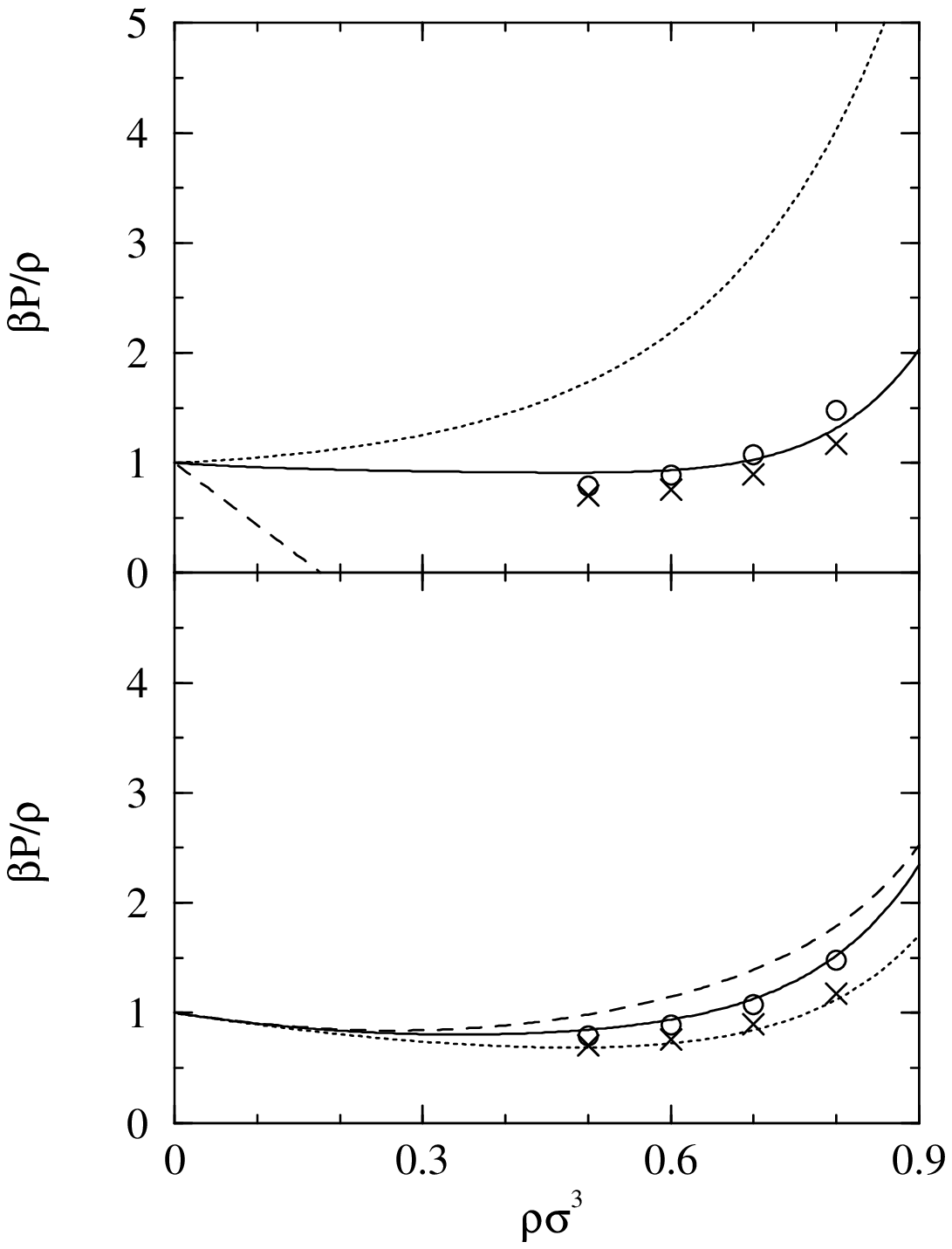}}
\caption{Same as figure~\protect\ref{fig:zcomp1} with $z=9\sigma^{-1}$ 
and $T^{*}=0.45$. MHNC and MC results are from Ref.~\protect\cite{costa}.}
\label{fig:zcomp3}
\end{figure}

In figures~\ref{fig:zcomp1}--\ref{fig:zcomp3} we report the results for
the equation of state of the hard-core Yukawa fluid. Reduced temperatures
$T^{*}=k_{\rm B}T/\epsilon$ and densities $\rho^{*}=\rho \sigma^{3}$ will 
be used throughout. We have considered both the linear and the non-linear
ORPA, and have compared them with the MHNC~\cite{costa} and Monte Carlo (MC) 
simulations~\cite{costa,shukla}. In order to determine the degree of 
thermodynamic consistency, results were obtained by all the three usual
routes to thermodynamics, namely the compressibility, the virial, 
and the energy route. We recall that these are different prescriptions 
to obtain the thermodynamics from the two-body correlations. Specifically,
in the compressibility route the key to the thermodynamics is the reduced
compressibility $\chi_{\rm red}$ determined as the structure factor $S(k)$ 
at zero wavevector $k$. In terms of the direct correlation function $c(r)$,
$\chi_{\rm red}$ is then given by
\begin{equation}
\frac{1}{\mbox{\hspace{0.2cm}}\chi_{\rm red}}=1-4\pi \rho 
\int_{0}^{\infty}\!\! dr \, r^{2} c(r) 
\, .
\label{comproute}
\end{equation} 
In the internal energy route, the starting point to determine the 
thermodynamics is instead the excess internal energy per particle $u$, 
obtained by the integral of the interaction weighted by the radial 
distribution function $g(r)$:
\begin{equation}
u=2\pi \rho \int_{\sigma}^{\infty}\!\!dr \, r^{2} g(r)w(r) \, .
\label{energyroute}
\end{equation} 
Finally in the virial route the thermodynamics is obtained from the pressure
$P$ as given by the virial theorem:
\begin{equation}
\frac{\beta P}{\rho}=1+\frac{2}{3}\pi \rho\left[\sigma^{3}g(\sigma^{+})-\beta
\int_{\sigma}^{\infty}\!\!dr \, r^{3}g(r)\frac{dw}{dr}(r)\right] \, ,
\label{virialroute}
\end{equation}  
where the contact value of the radial distribution function $g(\sigma^{+})$
appears because of the discontinuity of the hard-core plus tail interaction
considered here. While in an exact treatment Eqs.~(\ref{comproute}), 
(\ref{energyroute}), (\ref{virialroute}) are obviously bound to give 
equivalent results,
this is in general not true as soon as approximations are introduced. 
The extent to which the equations of state determined via the various
routes differ from one another is a useful test of the reliability 
of the theory. Figures~\ref{fig:zcomp1}--\ref{fig:zcomp3} show that 
the non-linear ORPA is much more consistent than the plain ORPA, especially
for short-ranged interactions. It is known~\cite{konior} 
that for the plain ORPA the internal energy route delivers the most reliable
results for the equation of state. This is true for the non-linear ORPA 
as well. The internal energy equations of state are actually very similar
in the two cases, although for short-ranged potentials and at low temperature
(see figure~\ref{fig:zcomp3}) the non-linear theory appears to be slightly 
superior. It is worthwhile observing that, while in the linear ORPA 
the pressure $P_{\rm en}$ obtained via the internal energy route always lies
between the compressibility pressure $P_{\rm comp}$ and the virial one 
$P_{\rm vir}$ with $P_{\rm vir}<P_{\rm en}<P_{\rm comp}$, this is not always 
the case with the non-linear ORPA. For instance, for an inverse interaction
range $z=9 \sigma^{-1}$ and reduced temperature $T^{*}=0.8$ 
the non-linear ORPA has again $P_{\rm vir}<P_{\rm en}<P_{\rm comp}$
(see figure~\ref{fig:zcomp2}), while for the same value of $z$ 
and $T^{*}=0.45$
it gives $P_{\rm comp}<P_{\rm en}<P_{\rm vir}$ (see figure~\ref{fig:zcomp3}).
For short-ranged interactions the energy-route equation of state is more 
accurate than the MHNC, which underestimates the simulation data. 

\begin{figure}
\centerline{\psfig{file=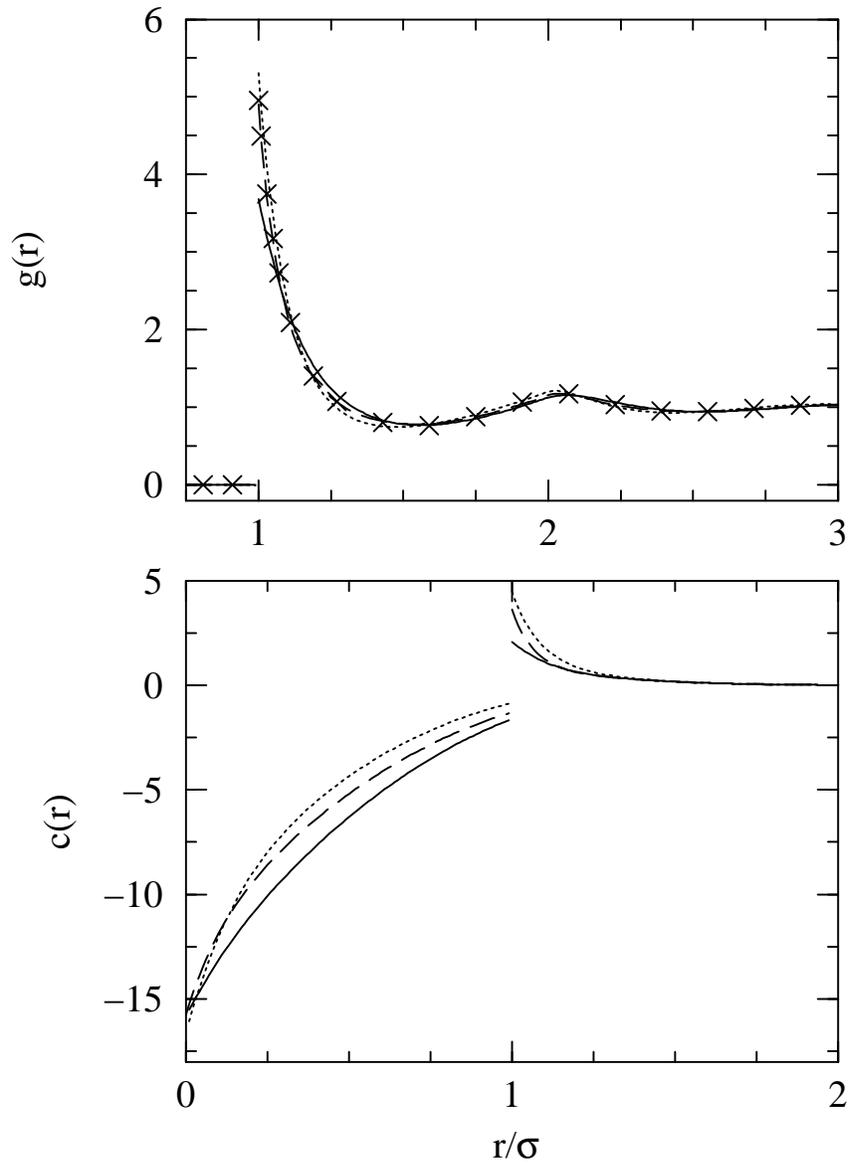}}  
\caption{Radial distribution function $g(r)$ (upper panel) 
and direct correlation function $c(r)$ (lower panel) 
of the hard-core Yukawa fluid with
$z=4\sigma^{-1}$ at reduced density $\rho^{*}=0.7$ and temperature
$T^*=0.62$, slightly above its critical value. Solid line, linear ORPA;
dotted line, non-linear ORPA; long-dashed line, MHNC; and crosses, MC data 
from Ref.~\protect\cite{costa}.}
\label{fig:gr1} 
\end{figure}

\begin{figure}
\centerline{\psfig{file=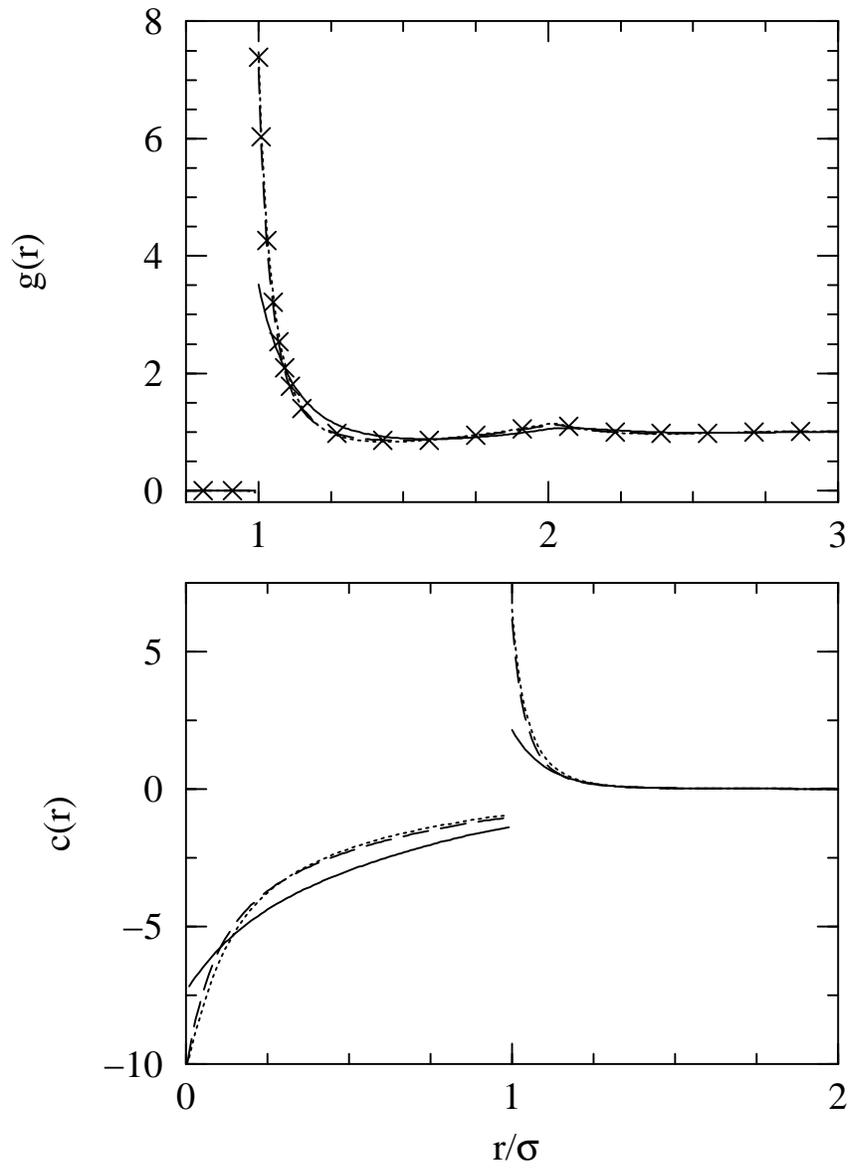}} 
\caption{Same as figure~\protect\ref{fig:gr1} for $z=9\sigma^{-1}$, 
$\rho^*=0.5$, $T^*=0.5$. According to simulation studies~\protect\cite{hagen}, 
this state point lies somewhat above the fluid-solid line.}
\label{fig:gr2}
\end{figure}

\begin{figure}
\centerline{\psfig{file=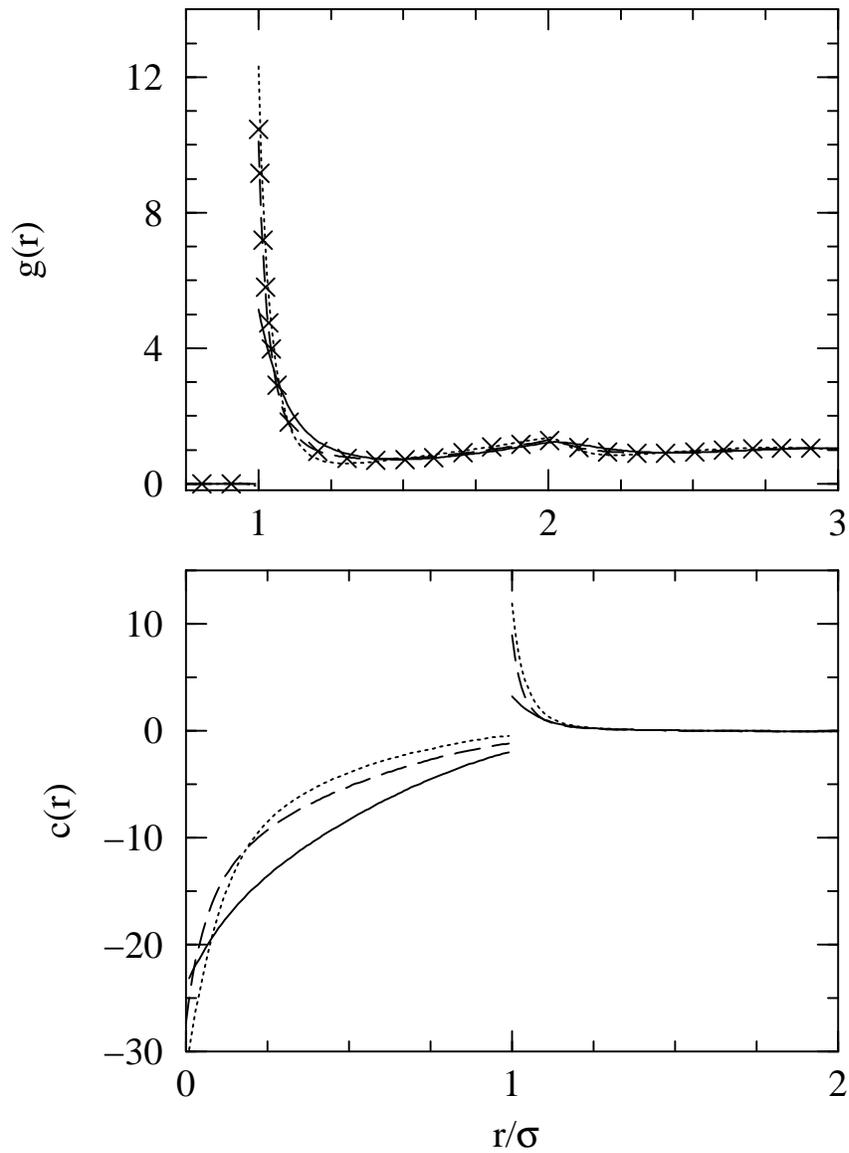}}
\caption{Same as figure~\protect\ref{fig:gr1} for $z=9\sigma^{-1}$, 
$\rho^*=0.8$, $T^*=0.4$. According to Ref.~\protect\cite{hagen}, 
this state point lies well inside the fluid-solid coexistence region.}
\label{fig:gr3}
\end{figure}

The radial distribution function $g(r)$ and the direct correlation function
$c(r)$ of the hard-core Yukawa fluid 
for several thermodynamic states and interaction ranges are shown 
in figures~\ref{fig:gr1}--\ref{fig:gr3}. Again, the results obtained 
by the linear and non-linear ORPA have been compared 
with the predictions of MHNC, which is known to describe two-body
correlations quite accurately. For $g(r)$, MC simulation data~\cite{costa} 
have also been reported. As anticipated, the plain ORPA appears to be
inadequate to describe the correlations for short-ranged interactions. 
In particular, the contact value of $g(r)$ is severely underestimated, 
with a relative error easily reaching $100\%$, and overall the profile 
of $g(r)$ does not show enough structure. The non-linear ORPA performs 
considerably better by comparison, although some 
discrepancies with the MC data emerge, especially at low temperature
and high density. These go in the opposite direction 
with respect to the linear ORPA, the contact value being overestimated,
and the structure being somewhat overemphasized. 
As a consequence, as shown in the figures, the exact $g(r)$ lies 
between the linear and non-linear ORPA results. Hence one expects 
that an accurate representation of the correlations can be obtained 
by performing a weighted average of the two closures. Such an approach 
has in fact been proposed by other authors~\cite{katsov} and referred 
to as tail interpolation method (TIM). For a hard-core plus tail potential, 
this consists in writing the contribution of the tail interaction 
to the direct correlation function outside the repulsive core as 
$c(r)-c_{\rm HS}(r)=\alpha [e^{-\beta w(r)}-1]+(1-\alpha)[-\beta w(r)]$,
where the parameter $\alpha$ is determined as a function of temperature 
and density by imposing consistency between the virial 
and the compressibility routes. In Ref.~\cite{katsov} this method was applied
to the hard-core Yukawa potential with $z=1.8 \sigma^{-1}$ as well as 
the Lennard-Jones potential via the Weeks-Chandler-Andersen~\cite{weeks} 
splitting of the interaction, and very accurate results for $g(r)$ 
were obtained. It could then be interesting to apply TIM also 
to narrower potentials. The possibility of improving the performance 
of the closure studied here by embedding it into a thermodynamically 
self-consistent scheme is discussed at the end of this section.  

\begin{figure}
\centerline{\psfig{file=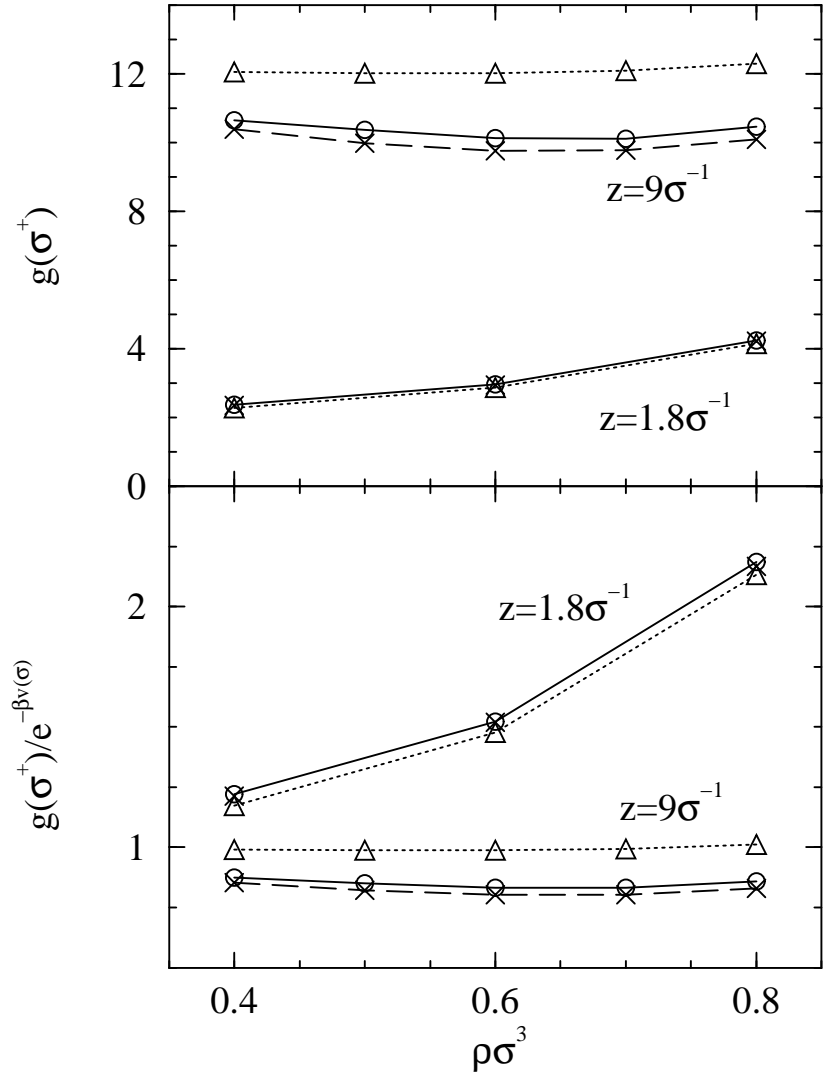}} 
\caption{Upper panel: contact value $g(\sigma^{+})$ 
of the radial distribution function of the hard-core Yukawa fluid 
with $z=1.8\sigma^{-1}$, $T^{*}=1.5$ and $z=9\sigma^{-1}$, $T^{*}=0.4$.  
Both isotherms lie slightly above the critical temperature 
for the corresponding $z$. Lower panel: contact value $g(\sigma^{+})$ 
rescaled by its low-density limit $e^{-\beta v(\sigma^{+})}$. 
Triangles, non-linear ORPA; crosses: MHNC; and 
circles, MC results~\protect\cite{costa}. The shorter-ranged interaction
has a much larger contact value, which remains nearly constant 
from $\rho^{*}=0.4$ to $\rho^{*}=0.8$ In the same interval, the contact value 
of the longer-ranged interaction increases by nearly a factor $2$ 
on increasing $\rho^{*}$.} 
\label{fig:contact}
\end{figure}

We observe that, unlike 
the linear ORPA, the non-linear closure~(\ref{orpanl}) becomes exact 
at low density $\rho$, as it yields the correct zeroth-order term
of the expansion of $c(r)$ in powers of $\rho$, thereby giving 
$c(r)\sim e^{-\beta v(r)}-1$, $g(r)\sim e^{-\beta v(r)}$ for 
$\rho \rightarrow 0$. This also implies that the second virial coefficient 
is correctly reproduced. Actually, it has been observed~\cite{louis} 
that for short-ranged interactions the form $g(r)\sim e^{-\beta v(r)}$ 
remains reasonably accurate near contact in a remarkably wide density interval. 
This is true in particular in the regime in which the fluid-solid
equilibrium curve broadens up and pre-empts the gas-liquid transition.
Figure~\ref{fig:contact} shows the contact values of $g(r)$ as predicted 
by non-linear ORPA, MHNC, and  MC simulations for the fluid with 
$z=9 \sigma^{-1}$ and $T^{*}=0.4$ as a function of density. 
According to simulation studies~\cite{hagen}, the states shown in the figure
are well inside the fluid-solid coexistence region. The contact value 
of $g(r)$ given by the non-linear ORPA is indeed very close to  
the value $e^{-1/T^{*}}\simeq 12.18$ for densities as high as $\rho^{*}=0.8$.
The insensitivity of the peak of $g(r)$ to the density is due 
to the fact that the clustering of the particles near contact is triggered
by the nearly sticky attractive interaction~\cite{louis}. This sharply 
contrasts the situation found in Lennard-Jones-like systems away from 
the critical point, where the structure is mainly determined 
by the excluded volume effect, hence being essentially that 
of a hard-sphere fluid at the same density. 

We now turn our attention to the critical point. We have obtained critical
temperatures and densities both for the hard-core Yukawa potential 
of Eq.~(\ref{yuk}) and the square-well potential of Eq.~(\ref{sq}). 
The results obtained for several interaction ranges by the compressibility 
and the internal energy route of the linear and non-linear ORPA have been 
compared with simulation 
results~\cite{lomakin,vega,elliott,orkoulas,lomba,hagen,scoza} 
in tables~\ref{tab:crityuk} and~\ref{tab:critsq}. 
The longest ranges considered correspond to the above-mentioned 
LJ-like values $z=1.8 \sigma^{-1}$, $\delta=0.5$ for the Yukawa 
and the square-well fluid respectively. 
The higher degree of thermodynamic consistency of the non-linear ORPA 
is evident from the better agreement between energy- 
and compressibility-route critical temperatures with respect to the linear
ORPA. In particular, the energy-route critical temperature of the non-linear
ORPA is in quite good agreement with the simulation results in the whole range 
interval considered, although a precise assessment of the accuracy 
of the theory in this respect is prevented by the rather large uncertainties
that usually affect critical point estimates based on simulations, 
as is apparent from the discrepancies between the simulation results  
reported in table~\ref{tab:critsq}.
The most reliable results are obtained by supplementing simulations 
with a finite-size scaling analysis; however, to our knowledge this 
was performed only 
for the longest ranges among those reported 
in the tables~\cite{scoza,brillantov,orkoulas}. It is clear
in any case that the compressibility route of linear ORPA gives the worst
predictions, by severely underestimating the critical temperature as the 
interaction range decreases. For the critical density $\rho_{c}$ 
the non-linear ORPA performs less accurately than for $T_{c}$: 
the compressibility route overestimates $\rho_{c}$ with respect to simulation,
while the energy route underestimates it. 
According to what observed above, for the square-well fluid 
the compressibility-route reduced critical temperatures $T^{*}_{c,{\rm l}}$ 
and $T^{*}_{c,{\rm nl}}$ predicted respectively by the linear 
and non-linear ORPA
are linked by the relation $1/T^{*}_{c,{\rm l}}=e^{1/T^{*}_{c,{\rm nl}}}-1$.
However, this relation does not hold for the critical temperatures given 
by the internal energy route. 

\begin{figure}
\centerline{\psfig{file=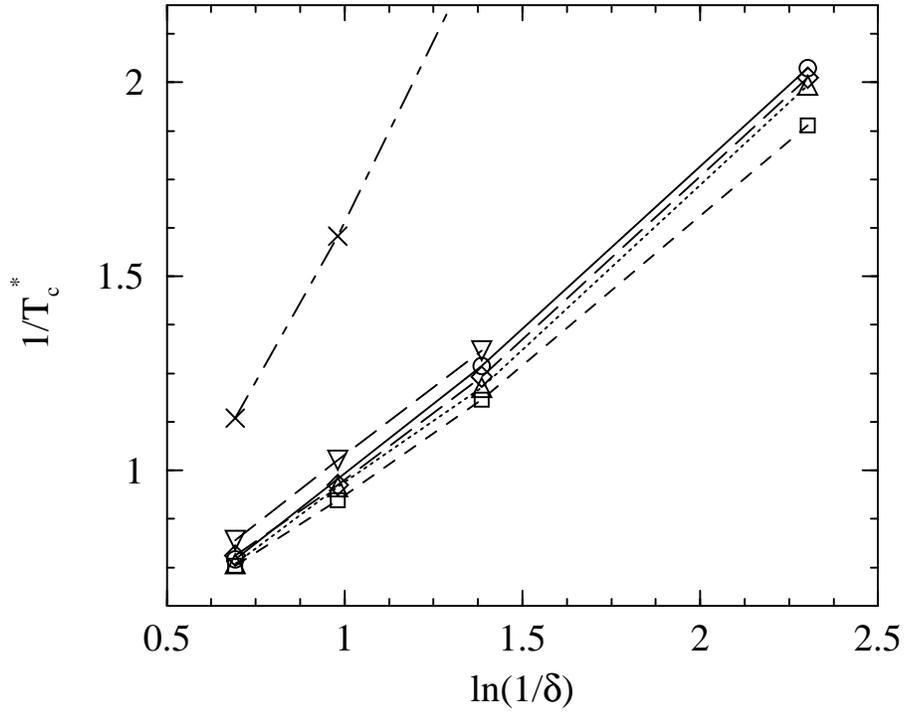}}
\caption{Inverse reduced critical temperature $1/T_{c}^{*}$
of the square-well fluid as a function of the width $\delta$ of the 
attractive well (see Eq.~(\protect\ref{sq})). 
The points
shown correspond to $\delta=0.5$, $\delta=0.375$, $\delta=0.25$, and
$\delta=0.1$. Crosses: linear ORPA compressibility route; squares, linear ORPA
energy route; triangles, non-linear ORPA compressibility route; diamonds,
non-linear ORPA energy route; circles, MC simulation data 
from Ref.~\protect\cite{lomakin}; and inverted triangles, MC simulation data
from Ref.~\protect\cite{vega}. Lines are a guide for the eye. 
The nearly linear dependence of $1/T_{c}^{*}$ 
on the logarithm of the inverse range $1/\delta$
stems from $B_{2}(T_{c})$ being a slowly varying function of $\delta$.}
\label{fig:virial}
\end{figure}

Since in introducing the closure~(\ref{orpanl}) we were inspired 
by the empirically observed behavior of the second virial coefficient 
at the critical temperature, it is worthwhile considering the results 
for $B_{2}(T_{c})$ given by the theory. These have also been reported 
in the tables. Moreover, in figure~\ref{fig:virial}   
we have plotted the reciprocal of the critical temperature
of the square-well fluid
as a function of the logarithm of the inverse well width $1/\delta$. 
If $B_{2}(T_{c})$ were rigorously independent of $\delta$, this plot 
would yield a straight line in the small-$\delta$ limit. The slow variation 
of $B_{2}(T_{c})$ with $\delta$ gives a nearly linear dependence of $1/T_{c}$
on $\ln(1/\delta)$ in a relatively wide interval of $\delta$, 
with the non-linear ORPA closely following the simulation results. 
However, neither
the simulation data nor the theory show evidence of $B_{2}(T_{c})$ saturating 
to a constant value in the interval considered. We observe that 
the linear ORPA gives rather poor results 
for $B_{2}(T_{c})$. In particular, the values obtained via the compressibility
route of linear ORPA are way too large in modulus and even show 
the wrong trend because of the strong underestimation of $T_{c}$.  

\begin{figure}
\centerline{\psfig{file=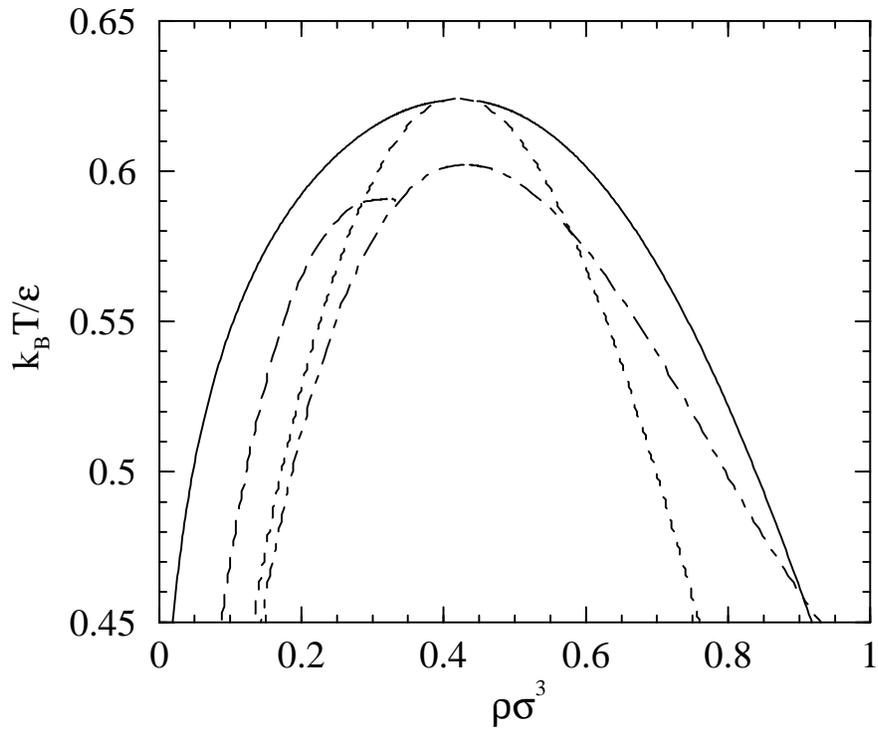}}
\caption{Dot-dashed line, spinodal curve of the hard-core Yukawa fluid with
inverse range $z=4\sigma^{-1}$ according to the compressibility route 
of non-linear ORPA. No solutions are found in the region bounded 
by this curve. Long-dashed line, locus of diverging compressibility 
according to the energy route of non-linear ORPA. Note that this
curve bumps into the spinodal. Dashed line, locus of diverging compressibility
according to the energy route of linear ORPA; solid line, coexistence
curve according to the energy route of linear ORPA.
The compressibility-route spinodal of linear ORPA lies well inside the 
coexistence curve and is located below the plane of the figure.}   
\label{fig:spinodal}
\end{figure}

\begin{figure}
\centerline{\psfig{file=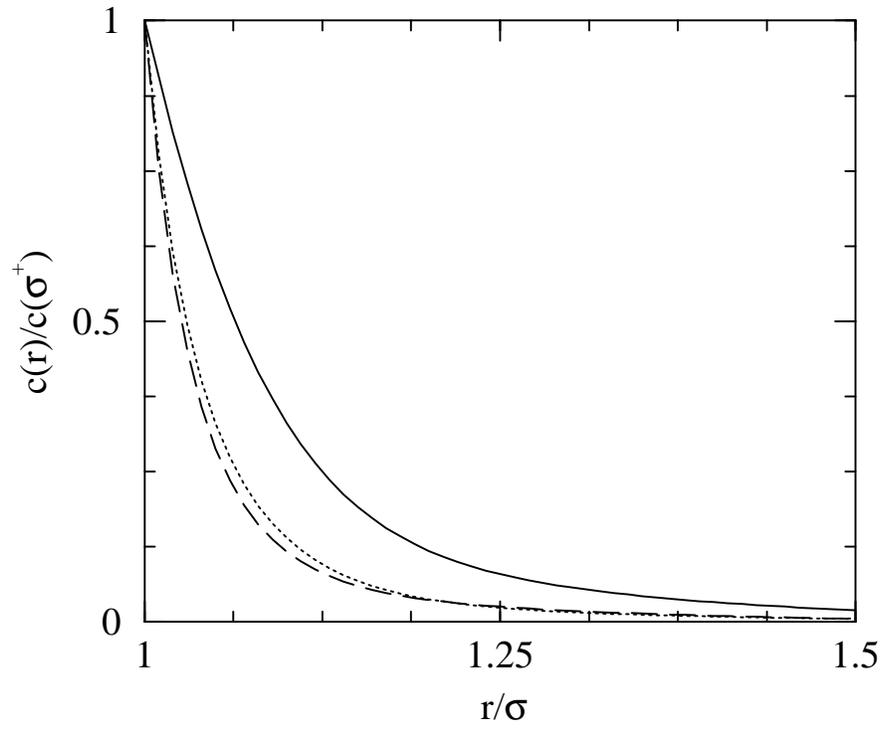}}  
\caption{Direct correlation function $c(r)$ of the hard-core Yukawa fluid 
with $z=9\sigma^{-1}$, $T^{*}=0.4$, $\rho^{*}=0.8$, as predicted 
by linear ORPA (solid line), non-linear ORPA (dotted line), 
and MHNC (long-dashed line), rescaled by its contact value $c(\sigma^{+})$.}  
\label{fig:direct}
\end{figure}

Below the critical temperature of the compressibility route, both the linear
and the non-linear ORPA give a spinodal curve, along which the compressibility
given by Eq.~(\ref{comproute}) diverges. Inside the region bounded by this
curve the approximation cannot be solved, as the integrand in the l.h.s.
of Eq.~(\ref{core2}) would develope a pole at non-zero wavevector 
and the integral would diverge. In order to obtain the coexistence curve, 
the problem arises of avoiding this forbidden region when determining 
the pressure and the chemical potential needed to impose the thermodynamic
equilibrium conditions. In the linear ORPA, this is achieved via 
the internal energy route, namely by getting the excess Helmholtz free energy
by integrating along each isochore the excess internal energy given 
by Eq.~(\ref{energyroute}). The success of this procedure depends 
on the fact that,
as shown in tables~\ref{tab:crityuk}, \ref{tab:critsq}, 
the critical temperature of the compressibility route of linear ORPA 
is considerably lower
than that of the internal energy route, at least for the simple model fluids
considered here. Therefore, the forbidden region lies well inside 
the internal energy coexistence boundary. However, we have found that 
this method is not effective when dealing with the non-linear ORPA. In fact,
as observed above, in this case the difference between the critical 
temperatures given by the compressibility and the internal energy route 
is relatively small. 
As a consequence, the relative location of the energy-route critical point 
and the spinodal curve is such that a substantial portion of the coexistence
curve should lie inside the forbidden region. This prevented us 
from determining the coexistence curve, save for a tiny part corresponding 
to the very top. The situation just described is illustrated 
in figure~\ref{fig:spinodal}, which shows the spinodal curve predicted 
by the non-linear ORPA for a hard-core Yukawa fluid with $z=4 \sigma^{-1}$,
together with the locus of points at which the internal-energy route 
compressibility diverges. For densities just above the critical one, 
the latter curve bumps into the spinodal. This rapidly widens 
as the temperature is lowered, so that also the high-density branch 
of the coexistence curve is swallowed by the forbidden region. Such a feature
is clearly quite a serious limitation of the non-linear ORPA, especially
since the thermodynamic consistency of the theory is not so high as to
allow one to reliably determine the pressure and the chemical potential
along a mixed integration path in the temperature-density plane, 
as it is often done with other integral-equation 
theories~\cite{caccamorev,caccamo}. A possible 
strategy to overcome it consists in using a functional form of $c(r)$ similar 
to that considered here in the context of a more sophisticated theory, 
such as the SCOZA or the HRT mentioned in the previous section. 
This could be accomplished by multiplying the Mayer function 
in Eqs.~(\ref{orpanl}), (\ref{orpanl2}) by a state-dependent amplitude 
$A(\rho, T)$. In the SCOZA $A(\rho, T)$ is determined by imposing consistency 
between the compressibility and the energy route, while in the HRT one 
requires the compressibility rule~(\ref{comproute}) to be satisfied along 
a process
in which the Helmholtz free energy of the system is obtained by gradually 
switching on the long-wavelength Fourier components of the interaction. 
Both of these procedures are expected to lead to a well-defined coexistence
curve. In fact, in the SCOZA the spinodal curve and the curve along which 
the internal-energy route compressibility diverges are forced to coincide. 
Since the coexistence boundary lies outside the latter curve, it will lie
outside the forbidden region as well. On the other hand the HRT, thanks 
to its renormalizaton-group structure, preserves the convexity of the free
energy for every thermodynamic state. No forbidden region appears, 
and the two-phase domain is correctly obtained as the locus of flat 
pressure-density isotherms~\cite{hrt}. As stated above, the accuracy 
of both the SCOZA
and the HRT for short-ranged interactions is affected by the 
functional form adopted for $c(r)$, similar to that of the linear ORPA. 
The predictions for $c(r)$ given by the linear and non-linear ORPA for
a number of different states of the Yukawa fluid have already been compared
to the accurate MHNC results in figures~\ref{fig:gr1}--\ref{fig:gr3}. 
In figure~\ref{fig:direct} the direct correlation function   
outside the repulsive core for the Yukawa fluid with $z=9 \sigma^{-1}$,
$T^{*}=0.4$, $\rho^{*}=0.8$ shown in figure~\ref{fig:gr3} 
has been rescaled by its value at contact $c(\sigma^{+})$. 
It appears that, even when both 
the linear and the non-linear ORPA differ considerably from the 
MHNC as for the state considered here,
the overall profile of $c(r)$ given by the non-linear ORPA can be adjusted
so as to reproduce that of the MHNC fairly well by suitably rescaling 
its amplitude, while this is not the case with the linear theory. 
Accordingly, we expect that implementing a closure similar to that 
of Eq.~(\ref{orpanl2}) within the HRT or the SCOZA scheme along the lines
sketched above will lead to an accurate description of thermodynamics,
correlations, {\em and} phase diagram of fluids with narrow attractive 
potentials. We intend to pursue this investigation in the future. 

\section{Conclusions}

We have considered a simple modification of the ORPA aimed at improving 
the performance of this theory for interactions which present a narrow
and deep attractive well. Its predictions have been compared with those 
of linear ORPA and MHNC as well as simulation data. The equation of state
is accurately reproduced, and the degree of thermodynamic consistency 
is considerably higher than in the plain ORPA. Also the two-body correlations
are more accurate than those of plain ORPA, although overall  
less accurate than in MHNC. The critical temperatures agree well 
with simulation results. In particular, the second virial coefficient 
at the critical temperature $B_{2}(T_{c})$ is found to depend rather weakly
on the interaction range. Below the critical temperature the spinodal curve
predicted by the theory can be easily obtained, but unfortunately 
most of the liquid-vapor coexistence
curve is not accessible, as it lies inside the region where no solutions
can be found. This drawback could be overcome by using a functional form
of the direct correlation function similar to that considered here 
in the context of a more sophisticated approach such as the HRT or the SCOZA. 
 
\section*{Acknowledgments}

We thank Dino Costa and Giuseppe Pellicane for making available to us the
files of their MC and MHNC results and their MHNC code.

\begin{table}
\begin{tabular}{l|rrr|rrr|rrr|rrr}  
\makebox[2cm]&
\multicolumn{3}{c|}{z=1.8} &
\multicolumn{3}{c|}{z=3} &
\multicolumn{3}{c|}{z=4} &
\multicolumn{3}{c}{z=7} \\
      &\multicolumn{1}{c}{$T_{c}^{*}$} & \multicolumn{1}{c}{$\rho_{c}^{*}$} &
\multicolumn{1}{r|}{$B_{2}(T_{c})$} & \multicolumn{1}{c}{$T_{c}^{*}$} &
\multicolumn{1}{c}{$\rho_{c}^{*}$} & \multicolumn{1}{r|}{$B_{2}(T_{c})$} &
\multicolumn{1}{c}{$T_{c}^{*}$} & \multicolumn{1}{c}{$\rho_{c}^{*}$} &
\multicolumn{1}{r|}{$B_{2}(T_{c})$} &
\multicolumn{1}{c}{$T_{c}^{*}$} & \multicolumn{1}{c}{$\rho_{c}^{*}$} &
\multicolumn{1}{r}{$B_{2}(T_{c})$} \\ \hline       
MC1 & 1.212 & 0.312 &$-5.90$ & & & & & & & & & \\ 
MC2 & 1.177 & 0.313 &$-6.24$ & 0.715 & 0.375
& $-6.14$ & 0.576 & 0.377 & $-5.92$ & & & \\  
MC3 & & & & & & & & & & 0.412 & 0.50 & $-5.51$ \\
ORPA (comp)   & 1.037 & 0.312 & $-7.89$ & 0.542 & 0.360 & $-11.10$
& 0.387 & 0.392 & $-15.35$& 0.213 & 0.460 & $-46.70$ \\
ORPA (energy) & 1.243 & 0.318 & $-5.61$ & 0.771 & 0.374 & $-5.16$
& 0.624 & 0.420 &$-4.75$ & 0.451 & 0.536 & $-4.05$ \\
ORPANL (comp) & 1.221 & 0.326 & $-5.82$ & 0.749 & 0.388 & $-5.52$
& 0.602 & 0.432 &$-5.29$ & 0.430 & 0.504 & $-4.77$ \\
ORPANL (energy) & 1.210 & 0.270 & $-5.92$ & 0.736 & 0.302 & $-5.75$
& 0.591 & 0.324 &$-5.55$ & 0.424 & 0.362 & $-4.98$ \\
\end{tabular} 
\caption{Critical constants of the hard-core Yukawa fluid. The reduced 
critical density $\rho_{c}^{*}$ and temperature $T_{c}^{*}$ 
from the compressibility and energy route of linear and non-linear ORPA
(indicated as ORPANL) 
are compared with simulation results. The second virial coefficient 
at the critical
temperature $B_{2}(T_{c})$ is in units of the hard-sphere volume 
$\pi\sigma^{3}/6$. MC1: Monte Carlo simulation data with finite-size scaling
from Ref.~\protect\cite{scoza}. MC2: Monte Carlo simulation data from 
Ref~\protect\cite{lomba}. MC3: Monte carlo simulation data 
from Ref.~\protect\cite{hagen}.}
\label{tab:crityuk}
\end{table}

\newpage

\begin{table}
\begin{tabular}{l|llr|llr|llr|llr}
\makebox[2cm]&
\multicolumn{3}{c|}{$\delta=0.5$} &
\multicolumn{3}{c|}{$\delta=0.375$} &
\multicolumn{3}{c|}{$\delta=0.25$} &
\multicolumn{3}{c}{$\delta=0.1$} \\ 
      &\multicolumn{1}{c}{$T_{c}^{*}$} & \multicolumn{1}{c}{$\rho_{c}^{*}$} & 
\multicolumn{1}{r|}{$B_{2}(T_{c})$} & \multicolumn{1}{c}{$T_{c}^{*}$} & 
\multicolumn{1}{c}{$\rho_{c}^{*}$} & \multicolumn{1}{r|}{$B_{2}(T_{c})$} &
\multicolumn{1}{c}{$T_{c}^{*}$} & \multicolumn{1}{c}{$\rho_{c}^{*}$} & 
\multicolumn{1}{r|}{$B_{2}(T_{c})$} &
\multicolumn{1}{c}{$T_{c}^{*}$} & \multicolumn{1}{c}{$\rho_{c}^{*}$} & 
\multicolumn{1}{r}{$B_{2}(T_{c})$}  
\\ \hline
MC1 & 1.218 & 0.310 & $-8.09$ &       &       &        &       &       & 
& & & \\
MC2 & 1.219 & 0.299 &$-8.08$ & 0.974 & 0.355 &$-7.46$ & 0.764 & 0.370 &$-6.30$& & & \\
MC3 & 1.299 & 0.322 &$-7.01$& & & & 0.788 & 0.392 &$-5.75$& 0.491 & 0.456 
&$-4.82$ \\ 
MD & 1.27 & 0.31 & $-7.38$ & 1.01 & 0.34 & $-6.82$ & 0.78 & 0.42 & $-5.93$ & & & \\  
ORPA (comp)   & 0.881 & 0.348 & $-16.05$ &0.624 & 0.388 &$-21.40$& 0.423 & 0.420
&$-32.73$& 0.158 & 0.537 & $-737.88$ \\
ORPA (energy) & 1.327 & 0.304 &$-6.68$ & 1.083 & 0.354 &$-5.72$ & 0.846 & 0.438 & $-4.61$ & 0.529 & 0.686 & $-3.43$ \\
ORPANL (comp) & 1.319 & 0.348 & $-6.78$ & 1.045 & 0.388 &$-6.26$ & 0.824 & 0.420 & $-5.01$ & 0.502 & 0.537 & $-4.38$ \\
ORPANL (energy) & 1.282 & 0.228 & $-7.22$ & 1.038 & 0.266 &$-6.36$ & 0.806 & 0.330 & $-5.37$ & 0.497 & 0.450 & $-4.57$ \\
\end{tabular}
\caption{Same as table~I for the square-well fluid. 
MC1: Monte Carlo simulation data with finite-size scaling 
from Ref.~\protect\cite{orkoulas}.  
MC2: Monte Carlo simulation data from Ref.~\protect\cite{vega}.
MC3: Monte Carlo simulation data from Ref.~\protect\cite{lomakin}. 
MD: molecular dynamics simulation data from Ref.~\protect\cite{elliott}.} 
\label{tab:critsq}
\end{table}

\newpage
\newpage

\end{document}